\documentclass[a4paper]{spie}  

\usepackage{amsmath,amsfonts,amssymb}
\usepackage{graphicx}
\usepackage[colorlinks=true, allcolors=blue]{hyperref}
\usepackage{color,soul}
\usepackage[utf8]{inputenc}
\usepackage[T1]{fontenc}
\usepackage{hyperref}

\hypersetup{
    colorlinks=true,
    linkcolor=blue,
    filecolor=magenta,      
    urlcolor=cyan,
    pdftitle={Overleaf Example},
    pdfpagemode=FullScreen,
    }
\title{ANDES, the high resolution spectrograph for the ELT: science goals, project overview and future developments}

\author[1,2]{A. Marconi}
\author[3]{M. Abreu}
\author[4,5]{V. Adibekyan}
\author[6]{V. Alberti}
\author[7]{S. Albrecht}
\author[8]{J. Alcaniz}
\author[9]{M. Aliverti}
\author[10,11]{C. Allende Prieto}
\author[12]{J. D. Alvarado Gómez}
\author[13,14]{C. S. Alves}
% ...
\affil[1]{Department of Physics and Astronomy, University of Florence, Italy}
\affil[2]{INAF - Osservatorio Astrofisico di Arcetri, Largo E. Fermi 5, I-50125 Firenze, Italy}
\affil[3]{Instituto de Astrof\'isica e Ci\^encias do Espa\c{c}o, Universidade de Lisboa, Faculdade de Ci\^encias, Campo Grande, PT1749-016 Lisboa, Portugal}
\affil[4]{Instituto de Astrof\'isica e Ci\^encias do Espa\c{c}o, Universidade do Porto, CAUP, Rua das Estrelas, PT4150-762 Porto, Portugal}
\affil[5]{Departamento de F\'isica e Astronomia, Faculdade de Ci\^encias, Universidade do Porto, Rua do Campo Alegre, 4169-007 Porto, Portugal}
\affil[6]{INAF - Osservatorio Astronomico di Trieste, via G. B. Tiepolo 11, 34143 Trieste, Italy}
\affil[7]{Stellar Astrophysics Centre, Department of Physics and Astronomy, Aarhus University, Ny Munkegade 120, 8000 Aarhus C, Denmark}
\affil[8]{Departamento de Astronomia, Observatório Nacional, 20921-400, Rio de Janeiro, RJ, Brazil}
\affil[9]{INAF - Osservatorio Astronomico di Brera, Via E. Bianchi 46, 23807 Merate (LC), Italy}
\affil[10]{Instituto de Astrof\'isica de Canarias (IAC), E-38200 La Laguna, Tenerife, Spain}
\affil[11]{Universidad de La Laguna, Dept. Astrof\'isica, E-38206 La Laguna, Tenerife, Spain}
\affil[12]{Leibniz Institute for Astrophysics Potsdam (AIP),  An der Sternwarte 16, D-14482 Potsdam, Germany}
\affil[13]{Centro de Astrof\'isica da Universidade do Porto, Rua das Estrelas, PT4150-762 Porto, Portugal}
\affil[14]{Department of Physics and Astronomy, University College London, Gower Street, London WC1E 6BT, United Kingdom}
% ...

\authorinfo{Send correspondence to A. Marconi (alessandro.marconi@inaf.it)}

\begin{document} 
\maketitle

\begin{abstract}
The first generation of ELT instruments includes an optical-infrared high resolution spectrograph, indicated as ELT-HIRES and recently christened ANDES (ArmazoNes high Dispersion Echelle Spectrograph).
ANDES consists of three fibre-fed spectrographs ([U]BV, RIZ, YJH) providing a spectral resolution of $\sim$100,000 with a minimum simultaneous wavelength coverage of 0.4-1.8 $\mu$m with the goal of extending it to 0.35-2.4 $\mu$m with the addition of an U arm to the BV spectrograph and a separate K band spectrograph. 
It operates both in seeing- and diffraction-limited conditions and the ﬁbre-feeding allows several, interchangeable observing modes including a single conjugated adaptive optics module and a small diﬀraction-limited integral field unit in the NIR. 
Modularity and fibre-feeding allows ANDES to be placed partly on the ELT Nasmyth platform and partly in the Coud\'e room.
ANDES has a wide range of groundbreaking science cases spanning nearly all areas of research in astrophysics and even fundamental physics. Among the top science cases there are the detection of biosignatures from exoplanet atmospheres, finding the fingerprints of the first generation of stars, tests on the stability of Nature's fundamental couplings, and the direct detection of the cosmic acceleration.
The ANDES project is carried forward by a large international consortium, composed of 35 Institutes from 13 countries, forming a team of almost 300 scientists and engineers which include the majority of the scientific and technical expertise in the field that can be found in ESO member states.
\end{abstract}


\section{Introduction}
\label{sec:intro}  

At first light in 2028, the European Extremely Large Telescope (ELT) will be the largest ground-based telescope at visible and infrared wavelengths. The flagship science cases supporting the successful ELT construction proposal were the detection of life signatures in Earth-like exoplanets and the direct detection of the cosmic expansion re-acceleration. It is no coincidence that both science cases require observations with a high-resolution spectrograph.
Over the past few decades high-resolution spectroscopy has been a truly interdisciplinary tool, which has enabled some of the most extraordinary discoveries spanning all fields of Astrophysics, from Exoplanets to Cosmology. Astronomical high-resolution spectrometers have allowed scientists to go beyond the classical domain of astrophysics and to address some of the fundamental questions of Physics. In the wide-ranging areas of research exploiting high-resolution spectroscopy, European scientists have been extremely successful, thanks to the exquisite suite of medium/high-resolution spectrographs that ESO provides to its community. UVES, FLAMES, CRIRES, X-Shooter and HARPS have enabled European teams to lead in many areas of research. ESPRESSO, which has recently joined this suite of very successful instruments, is fulfilling its promise of truly revolutionizing some of these research areas. The scientific interest and high productivity in this field of research is reflected by the fact that more than 30\% of ESO publications can be attributed to medium/high-resolution spectrographs.
However, it is becoming increasingly clear that, in most areas of research, high-resolution spectroscopy has reached or is approaching the \textit{photon-starved} regime at 8-10m class telescopes. Despite major progress on the instrumentation front, further major advances in these fields desperately require the collecting area of Extremely Large Telescopes.
When defining the ELT instrumentation, ESO commissioned two phase-A studies for high-resolution spectrographs, CODEX (Pasquini et al.\cite{pasquini:2010}, covering the 370 nm – 710 nm wavelengths range) and SIMPLE (Origlia et al.\cite{origlia:2010}, covering the 840 nm – 2400 nm wavelengths range), which were started in 2007 and completed in 2010. These studies demonstrated the importance of optical and near-IR high-resolution spectroscopy at the ELT and ESO thus decided to include a High-REsolution Spectrograph (HIRES) in the ELT instrumentation roadmap. 
Soon after conclusion of the respective phase A studies the CODEX and SIMPLE consortia realized the great scientific importance of covering the optical and near-infrared spectral ranges simultaneously. This marked the birth of the HIRES initiative (\href{https://hires-eelt.org}{hires-eelt.org}) that started developing the concept of an X-Shooter-like spectrograph, but with high resolution, capable of providing R$\sim$100.000 over the full UV, optical and near infrared wavelength ranges. Following a community workshop in September 2012 the HIRES Initiative prepared a White Paper summarizing a wide range of science cases proposed by the community (Maiolino et al.\cite{maiolino:2013}), and a Blue Book with a preliminary technical instrument concept (Riva et al. 2015). 

With the start of construction of the ELT, the HIRES Initiative has decided to organize itself as the HIRES Consortium and has recruited additional institutes, which expressed their interest in HIRES. The consortium, strongly motivated by the unprecedented scientific achievements that the combination of such an instrument with the ELT will enable, was commissioned to perform a Phase A study by ESO which was successfully carried out in 2016-2018.
The Consortium was then involved in several pre-Phase B activities in preparation of the start of construction until the ESO Council approved the Construction of HIRES in December 2021. 
The instrument was then renamed ANDES (ArmazoNes high Dispersion Echelle Spectrograph, \href{https://andes.inaf.it}{andes.inaf.it}) and the ANDES Consortium started to formally organize with the signature of the ANDES Consortium agreement where 25 Partners represent 33 institutes from ESO Member States, Brazil, Canada and USA.
ANDES started Phase B activities in September 2022: this Phase B1 was successfully concluded in October 2023 with the System Architecture Review (SAR). The construction has finally formally started in June 2024, with the signature of the Construction Agreement with ESO. The conclusion of Phase B and the beginning of subsequent phases is planned in mid-2025, with the aim of bringing ANDES to the telescope in late 2031/early 2032.

This paper provides a general description of the ANDES project, science and consortium.
In section 2 we describe the ANDES science goals and priorities, in section 3 the instrument concept, in section 4 the consortium and its organization, and in Section 5 cost estimates and schedule.

\section{Science Goals}

\subsection{Exoplanets and Protoplanetary Disks}
\begin{figure}[ht]
\begin{center}
\begin{tabular}{c} 
   \includegraphics[width=0.9\linewidth]{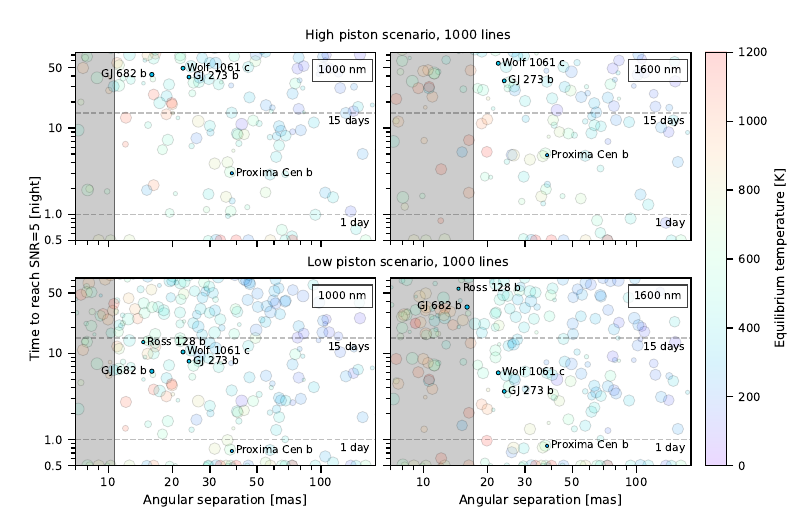}
\end{tabular}
\end{center}
\caption{ \label{fig:exo_acessible}
Angular separation between the star and the planet as a function of the exposure time (in nights) necessary to reach a SNR=5 for the known population of planets accessible for atmospheric characterization through reflected light detection at 2 selected wavelengths, we assumed a low piston scenario for the AO correction, and a cross correlation over 1000 absorption lines in the planetary atmosphere. The grey area represents angular separations smaller than twice the ratio between wavelength and telescope diameter. Adapted from Palle et al.\cite{palle:2024}.}
\end{figure}

\textbf{Characterization of exoplanets atmospheric composition and the exploration of habitable zone planets.}  
The study of exoplanet atmospheres is essential for understanding the diversity, formation, and potential habitability of planetary systems. Atmospheric analysis can reveal a planet's composition, structure, and climate, providing insights into its origins and evolutionary history. Specifically, the detection of atmospheric gases such as oxygen, methane, and water vapor can indicate biological activity, making these studies crucial in the search for extraterrestrial life.
Current techniques for studying exoplanet atmospheres include transmission and emission spectroscopy. During a transit, starlight passes through the planet's atmosphere, allowing for the detection of specific molecular signatures in the transmission spectrum. Emission spectroscopy measures the thermal emission from a planet, offering details about its atmospheric composition and temperature. The combination of ELT and ANDES promises to significantly advance the field of exoplanet atmospheric research allowing to pursue ambitious scientific objectives like the characterization of exoplanets atmospheric composition and the exploration of habitable zone planets. 

Understanding the chemical composition of exoplanet atmospheres is crucial for identifying potentially habitable planets. ANDES's high spectral resolution (R $\simeq$ 100,000) will enable the detection of faint spectral lines, in between telluric absorption lines, facilitating the identification of trace gases and isotopic ratios. This enhanced sensitivity is essential for detecting biomarkers such as oxygen and methane, which are indicative of biological processes. 
A comprehensive wavelength coverage (0.5-1.8 $\mu$m, with a goal to extend to 0.38-2.4 $\mu$m) is also vital for a full characterization of exoplanet atmospheres: ANDES will observe key molecular bands, including those of H$_2$O, O$_2$, CO$_2$, CH$_4$, NH$_3$, and other species. This broad coverage ensures that all significant atmospheric components can be detected and analyzed, providing a complete picture of atmospheric composition.
Detailed atmospheric dynamics, such as weather patterns and wind speeds, is essential for characterizing exoplanet climates. ANDES's high spectral resolution will allow for precise measurements of Doppler shifts in spectral lines, providing detailed information on atmospheric circulation and dynamics.

While exoplanets atmospheres can be studied with transmission spectroscopy, high-contrast observations can provide direct imaging of exoplanets, allowing to minimize contamination from stellar light. ANDES's ability to achieve high-contrast observations will enable direct imaging of exoplanets, particularly those that do not transit their host stars. This capability is essential for studying reflected light from exoplanets and for investigating planets in wider orbits that are not accessible through transit spectroscopy.

One of the primary scientific goals of ANDES is to characterize the atmospheres of small, rocky exoplanets in the habitable zone. These planets are of particular interest because they may have conditions suitable for liquid water and, potentially, life. Detailed studies of systems like Proxima Cen and TRAPPIST-1, which contains several Earth-sized planets in the habitable zone, will provide valuable information on the atmospheric conditions of potentially habitable worlds. For instance, a putative CO$_2$-dominated atmosphere on TRAPPIST-1b planet can be detected and studied at $>12\sigma$ in 10 transits with transmission spectroscopy (the effects of stellar contamination can be disentangled at high spectral resolution), while Proxima Cen b can be detected at $>5\sigma$ in 7 h with direct imaging in reflection. The detection of O$_2$ and H$_2$O in Proxima Cen b at $>5\sigma$ will require instead 60 and 4 nights respectively, but these numbers can vary widely depending on the bulk composition of the atmosphere.
Figure \ref{fig:exo_acessible} presents the currently known population of exoplanets accessible for atmospheric characterization through AO-assisted reflected light detection at different wavelengths while Table \ref{tab:golden} shows the Golden Sample of planets for ANDES, i.e., the currently known five rocky exoplanets orbiting in the habitable zone of their host stars most favorable for ANDES atmospheric characterization.
\begin{table}[ht]
\begin{center}
\begin{tabular}{c} 
   \includegraphics[width=0.9\linewidth]{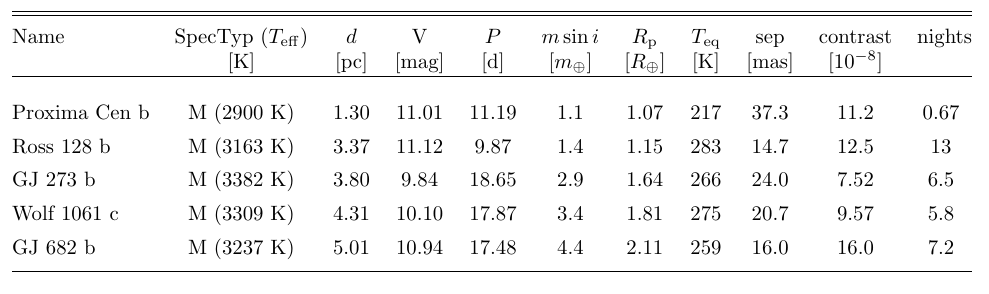}
\end{tabular}
\end{center}
\caption{ \label{tab:golden} The Golden Sample planets for ANDES, i.e., the currently known five rocky planets orbiting in the  habitable zone of their host stars most favorable for ANDES atmospheric chracterization. Along with several planets properties, we tabulate the number of nights necessary for detecting each planet's atmosphere.}
\end{table}

Finally, ANDES will work in conjunction with other major observatories, such as NASA-ESA's JWST and ESA's ARIEL mission. Combining high-resolution ground-based data from ANDES with space-based observations will enhance the overall scientific output. This synergy is particularly important for cross-verifying results and obtaining a more comprehensive understanding of exoplanet atmospheres across different spectral ranges.

\textbf{Protoplanetary disks and the formation and evolution of planetary systems.}
Understanding the formation and evolution of planetary systems necessitates a detailed study of protoplanetary disks. These disks are the birthplaces of planets, containing the gas and dust from which planetary bodies coalesce. Investigating the composition and distribution of gas within these disks, particularly in the inner regions (within 20 astronomical units), is crucial for comprehending the processes that lead to planet formation and migration.

Protoplanetary disks are best studied through high-resolution spectroscopy, which allows for the separation and analysis of different gas components and mechanisms at play. ANDES, coupled with the large collecting area of the ELT, and with its high spectral resolution and spatial capabilities, will be instrumental in characterizing the gas properties in these regions. Key scientific objectives include  the characterization of inner disk gas, the study of disk evolution and the investigation of disk winds and gas dispersal.

By studying forbidden lines of atomic and weakly ionized species at low radial velocities, ANDES will trace the gas in the inner disk regions. This will enable the differentiation of bound gas from disk winds, allowing for the derivation of physical parameters such as density, temperature, and ionization fraction. Observations of circumstellar disks around young stellar objects of various ages will help trace the evolution of disks from protostellar stages (around 100,000 years old) to more evolved phases (up to 10 million years old). This will shed light on how disks dissipate and how planetary systems form and evolve. ANDES will examine the different mechanisms contributing to gas dispersal in the inner disk, such as magnetospheric accretion, jets, and disk winds. High-resolution data will help distinguish these mechanisms and quantify their impact on disk evolution.
The data obtained from ANDES will complement observations from mid-infrared instruments like JWST/MIRI and ELT/METIS, providing a comprehensive understanding of protoplanetary disk dynamics and their role in planet formation. 

A detailed description of the science objectives in the field of exoplanets and protoplanetary disk is provided in the white paper by Palle at al.\cite{palle:2024}.

\subsection{Stars and Stellar Populations}
\begin{figure}[ht]
\begin{center}
\begin{tabular}{c} 
   \includegraphics[width=0.7\linewidth]{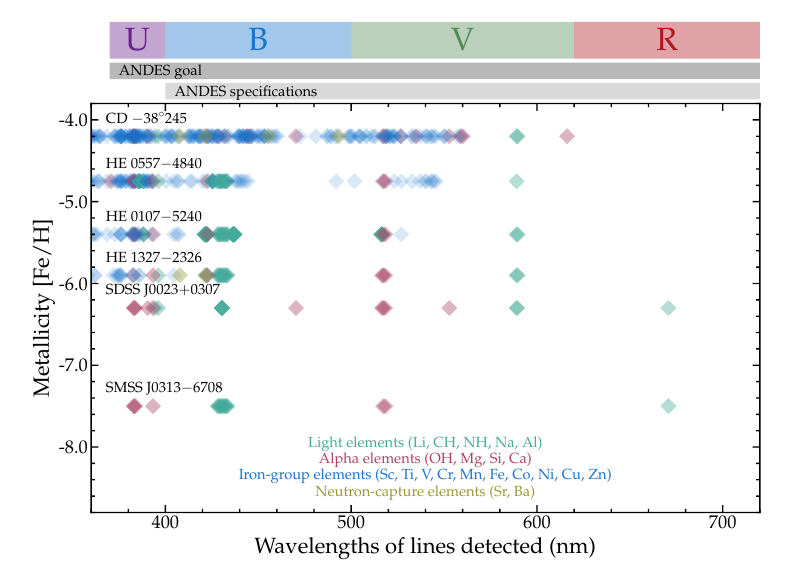}
\end{tabular}
\end{center}
\caption{ \label{fig:primitive} 
Wavelengths of absorption lines detected in the optical spectra of six stars known at present with [Fe/H] $<-4$. 
Each point represents one line in one star, and different elements are marked by different colors. The design of ANDES will enable the detection of the vast majority of these lines, covering all major nucleosynthesis groups, in many more stars that retain the fossil records of metals produced by the first stars. Figure from Roederer et al.\cite{roederer:2024}
}
\end{figure}

The study of stars and stellar populations is fundamental to understanding the formation and evolution of galaxies. Stars are the primary building blocks of galaxies and serve as the key drivers of chemical enrichment and energy distribution within these systems. By investigating the properties of stars in various evolutionary stages and environments, astronomers can reconstruct the history of star formation and the processes that govern stellar evolution. This knowledge is crucial for interpreting observations of distant galaxies and for refining theoretical models of galaxy formation and evolution.
Key scientific objectives for ANDES in the study of stars and stellar populations are the following.

\textbf{First Stars.}
The first stars, or Population III (Pop III) stars, are believed to have formed from metal-free gas at redshifts greater than 15-30. These stars played a crucial role in ending the cosmic Dark Ages by producing the first ionizing photons, supernovae, metals, and stellar-mass black holes. Despite their significance, Pop III stars have not been directly observed. ANDES aims to reveal the nature and end states of these stars by searching for long-lived, low-mass metal-free Pop III stars. The high spectral resolution (R $\sim$ 100,000) of ANDES is critical for detecting the faint spectral lines of hydrogen without contamination from metal lines (see, e.g., Fig. \ref{fig:primitive}). By collecting spectra of candidate metal-free stars with [Fe/H] $< -4$, ANDES can provide stronger constraints on the characteristic mass and low-mass end of the Pop III initial mass function (IMF).
\begin{figure}[ht]
\begin{center}
\begin{tabular}{c} 
   \includegraphics[width=0.6\linewidth]{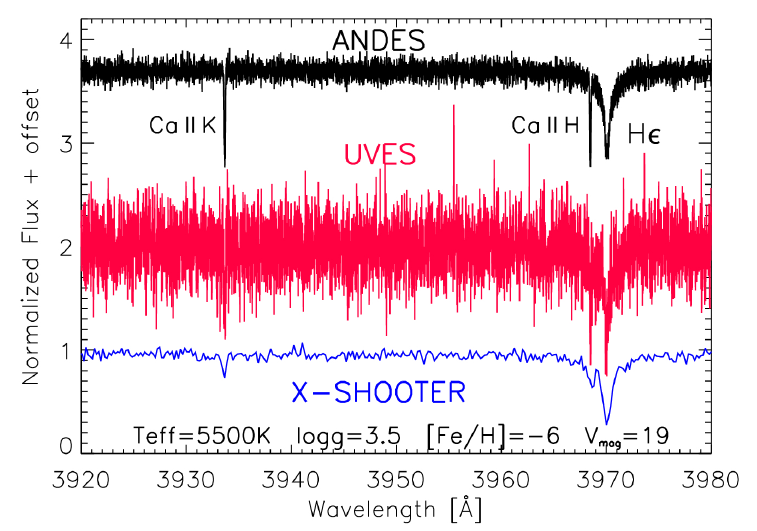}
\end{tabular}
\end{center}
\caption{ \label{fig:subgiant} 
A simulated 1-hour spectrum of a subgiant second-generation star with a total metallicity of 10$^{ -6}$ times the Solar metallicity in the U channel around Ca II H \& K lines. Random noise has been added to simulate the S/N for ANDES (black), UVES (red), and X-SHOOTER (blue). Figure from Roederer et al.\cite{roederer:2024}.
}
\end{figure}

\textbf{Stellar Populations in the Galactic Bulge.}
Understanding the age distribution of stars in the Galactic bulge is essential for deciphering the formation and evolution processes of this region. Observational evidence on this topic is conflicting, with some studies suggesting the bulge is almost entirely old, while others report a significant fraction of young and intermediate-age stars. ANDES, with its large collecting area of the ELT, will obtain high-resolution spectra of turnoff and subgiant stars in the bulge, even without gravitational microlensing magnification. These spectra will enable the determination of parameters, metallicities, and detailed chemical abundances, allowing for stellar age inference with about 25\% precision (Fig. \ref{fig:subgiant}). The broad wavelength coverage (0.40 to 1.80 $\mu$m) and high signal-to-noise ratios provided by ANDES are crucial for identifying and analyzing the numerous spectral lines necessary for these determinations. These data will help map the age distribution across the bulge and provide novel constraints on its formation and connections to other stellar populations.

\textbf{Stellar Populations in Dwarf Galaxies.}
The Local Group's dwarf galaxies offer unique insights into hierarchical mass assembly. These galaxies, spanning a wide range of stellar masses and chemical compositions, are prime targets for studying the imprints of the first stars and the earliest nucleosynthesis events. ANDES will enable high-resolution and high-signal-to-noise optical spectroscopy of stars in these galaxies, which are often too faint for current telescopes. This capability will allow for detailed chemical compositions to be derived, providing insights into the progenitors of supernovae, the production of heavy elements, and the origins of extreme carbon enhancement observed in many metal-poor stars. The adaptive optics integration with ANDES will enhance its capability to obtain high-quality spectra of these faint stars, making it possible to study their detailed compositions and evolutionary histories.

In summary, the capabilities of ANDES are integral to advancing the study of stellar populations. By providing high-resolution spectra, broad wavelength coverage, and the ability to observe faint and distant stars, ANDES will significantly enhance our understanding of the formation and evolution of the first stars, the Galactic bulge, and dwarf galaxies in the Local Group. These studies will provide critical insights into the processes governing stellar evolution and the formation of galaxies.

A detailed description of the science objectives for stars and stellar populations is provided in the white paper by Roederer et al.\cite{roederer:2024}.

\subsection{Galaxy Formation and evolution and the intergalactic medium}

The study of the reionisation epoch (EoR) is crucial for understanding the formation and evolution of the universe's first structures. This epoch marks the transition from a neutral to an ionized intergalactic medium (IGM) to the emergence of the first luminous sources. Absorption spectroscopy of high-redshift quasars and gamma-ray bursts (GRBs) provides a direct probe into the conditions of the IGM during this transformative period.
ANDES will be able to addresse several important science goals detailed below.

\begin{figure}[ht]
\begin{center}
\begin{tabular}{c} 
   \includegraphics[width=0.7\linewidth]{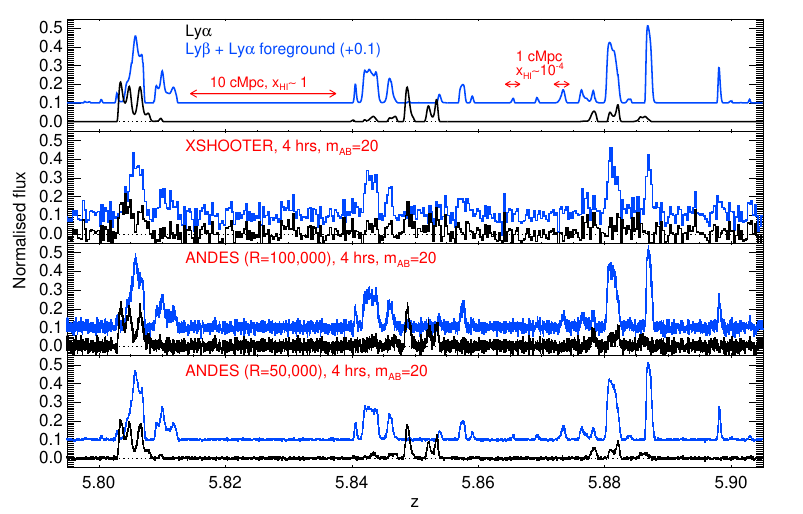}
\end{tabular}
\end{center}
\caption{ \label{fig:forests} 
The upper panel shows a segment of the Lyman-$\alpha$ (black curve) and coeval Lyman-$\beta$ (blue curve) forests in the spectrum of a z = 6.1 quasar, drawn from a cosmological hydrodynamical simulation coupled with radiative transfer. Note that the Lyman-$\beta$ forest flux has been offset by +0.1 for presentation purposes, and it also includes foreground Lyman-$\alpha$ forest absorption. The lower three panels show simulated observations with an exposure time of 4 hours assuming a magnitude $m_{AB} = 20$ for the target quasar, obtained with (from top to bottom): VLT-XSHOOTER at R$ \sim$ 10,900) with S/N$\sim$12, ELT-ANDES for the baseline resolution of R = 100,000, reaching S/N$\sim$80 and for the lower resolution mode with R = 50,000 with S/N$\sim$110. All signal-to-noise ratios are per 10 km/s spectral element. Figure from D'Odorico at al.\cite{dodorico:2024}.
}
\end{figure}

\textbf{Characterising the Epoch of Reionisation.}
The high spectral resolution (R $\sim$ 100,000) of ANDES will enable precise measurements of the Lyman-$\alpha$ and Lyman-$\beta$ forests in the spectra of $z \ge 6$ quasars and GRBs. By resolving these lines, ANDES will map the distribution of neutral and ionised hydrogen in the IGM, allowing us to track the progression of reionisation (Fig. \ref{fig:forests}). Detailed studies of these forests will reveal the size and distribution of ionised bubbles and neutral patches, shedding light on the sources driving reionisation, such as the first stars, galaxies and active galactic nuclei (AGNs). The unprecedented sensitivity of ANDES, owing to the ELT's large collecting area, will allow us to observe fainter quasars and GRBs, increasing the number of targets and providing a statistically significant sample strengthening constraints on the size and distribution of ionised and neutral regions, offering a more nuanced picture of how reionisation progressed spatially across the universe.

\begin{figure}[ht]
\begin{center}
\begin{tabular}{c} 
   \includegraphics[width=0.8\linewidth]{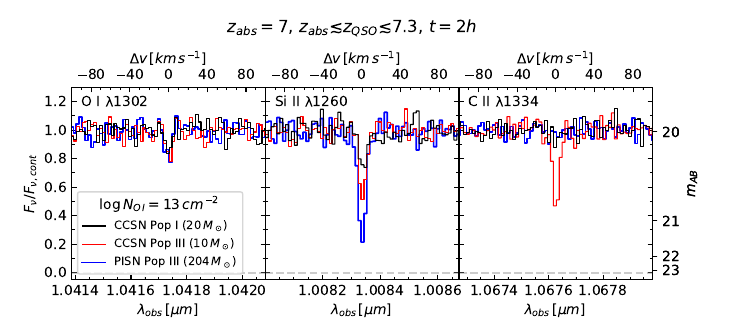}
\end{tabular}
\end{center}
\caption{ \label{fig:metal_abs} 
Simulated ANDES spectrum of a metal absorption system at $z = 7.0$, with N(O I) $= 10^{13} cm^{-2}$ along the line of sight to a background $m_{AB} = 20$ quasar observed for 2 hours. The panels display zoomed-in regions around the OI $\lambda$1302, SiII $\lambda$1260 and CII $\lambda$1334 absorption features in three different scenarios enriched by the ejecta from: the core collapse supernova (CCSN) of a $Z = 1/3 Z_\odot$, $m_\star = 20 \,M_\odot$ star (black line), the Pop III CCSN of a $m_\star = 10 \,M_\odot$ star (red line) and the Pop III PISN from a $m_\star = 204 \,M_\odot$ star (blue line). The three scenarios can be clearly distinguished from the ANDES spectrum. Figure from D'Odorico at al.\cite{dodorico:2024}.
}
\end{figure}

\textbf{Chemical Enrichment of the Early Universe.}
Above redshift z $\sim$ 5.5, the Lyman forest becomes less informative due to complete absorption of the quasar flux. However, ANDES's ability to detect metal absorption lines in the near-IR spectrum will allow studies of the the IGM physical and chemical properties even when hydrogen lines are saturated. 
ANDES's broad wavelength coverage (0.4-1.8 $\mu$m, with a goal of 0.35-2.4 $\mu$m) will allow simultaneous observation of multiple absorption lines from various elements such as carbon, oxygen, silicon and iron within the epoch of Reionization. In particular, the transition OI at 1302 A is a tracer of neutral gas  and can provide a complementary probe to the Lyman-series lines to study the ionization state of the IGM. Additionally, the measurement of the detailed chemical abundances in dense neutral environments will be crucial to detect the signatures of the enrichment by the first generation of stars (Fig. 5). In this respect, ANDES will allow for the first time to measure critical trace elements such as Zinc (Fig. \ref{fig:metal_abs}). By detecting and analyzing these metal lines, ANDES will provide insights into the production and distribution of heavy elements in the early universe, helping us understand the contribution of the first supernovae and the initial stages of galaxy formation.

\textbf{Understanding the Relationship Between Galaxies and the IGM.}
ANDES will also facilitate the study of the relationship between galaxies and the IGM by correlating absorption features with galaxies detected in emission through other instruments like JWST. This multi-wavelength approach will help us understand how galaxies influenced their surroundings and contributed to the reionisation process. By combining absorption and emission data, ANDES will provide a comprehensive view of the interactions between galaxies and the IGM, offering insights into the feedback mechanisms that regulated star formation and galaxy growth.
The fine spectral resolution of ANDES will enable precise measurements of the thermal broadening of absorption lines, allowing to infer the temperature of the IGM during the reionisation epoch. Understanding the thermal state of the IGM is essential for constraining the sources of heating, such as UV radiation from early galaxies and quasars. The thermal history provides critical information about the energy budget of the universe and the efficiency of ionising photon production. ANDES's ability to resolve these thermal features will help refine models of IGM heating and cooling processes.

\textbf{Extragalactic transients.}
Extragalactic transients, such as gamma-ray bursts (GRBs), superluminous supernovae (SLSNe), and novae, provide crucial insights into the physical and chemical properties of their host galaxies. These short-lived events are often faint, posing significant observational challenges that  ANDES  aims to overcome with its unprecedented sensitivity and near-infrared capabilities.
GRBs, associated with massive star explosions, serve as cosmic beacons to study the interstellar medium (ISM) of distant galaxies. SLSNe, being 10 to 100 times more luminous than typical supernovae, can be observed at high redshifts, shedding light on the properties of low-metallicity host galaxies. The exact mechanisms behind these explosions remain debated, with potential links to very massive stars or Population III stars. Classical novae involve thermonuclear explosions on white dwarfs and provide insights into the synthesis of heavy elements like lithium.
ANDES's high resolution and wide spectral coverage will enable detailed studies of these transients, enhancing our understanding of the early universe and the lifecycle of galaxies. Rapid response modes will be particularly beneficial for capturing the transient phases of GRBs, ensuring timely and detailed observations.

In summary, the capabilities of ANDES will allow for unprecedented studies of the reionisation epoch, providing detailed maps of the ionisation state, chemical enrichment, and thermal history of the early universe. These studies are essential for understanding the formation and evolution of the first cosmic structures and the processes that shaped the universe as we see it today.

A detailed description of the science objectives for galaxy formation and evolution, and the intergalactic medium is provided in the white paper by D'Odorico at al.\cite{dodorico:2024}.

\subsection{Cosmology and Fundamental Physics}

Spectroscopy has historically been a catalyst for significant advancements in fundamental physics. In the 21st century, ANDES at ESO’s ELT is poised to continue this tradition, offering unprecedented opportunities to probe the fundamental aspects of the universe. By leveraging high-resolution spectroscopy, ANDES aims to address critical questions in cosmology and fundamental physics. Here we outline the primary scientific goals and the key capabilities of ANDES that will enable these investigations.

\textbf{Big Bang Nucleosynthesis (BBN).}
Big Bang Nucleosynthesis provides a critical probe of the early universe by measuring the primordial abundances of light elements such as deuterium (D), helium-3 ($^3$He), helium-4 ($^4$He), and lithium-7 ($^7$Li). These abundances are sensitive to the baryon density, the expansion rate, and the particle content of the universe, thus offering insights into potential deviations from the Standard Model of particle physics and cosmology.
ANDES's high spectral resolution (R$\sim$100,000) will allow for the precise measurement of the primordial ratios of D/H, $^3$He/$^4$He, and $^7$Li/H in quasar absorption line systems. The large collecting area of the ELT will facilitate observations of faint, distant objects necessary for accurate abundance measurements, testing the consistency of the cosmological model and the Standard Model.
By combining measurements of multiple primordial elements with theoretical BBN calculations, ANDES aims to uncover any deviations from the Standard Model. The wide wavelength coverage (0.35–2.4 $\mu$m) is crucial for capturing the necessary spectral lines across various redshifts, providing a comprehensive test for new physics.
\begin{figure}[ht]
\begin{center}
\begin{tabular}{c} 
   \includegraphics[width=0.8\linewidth]{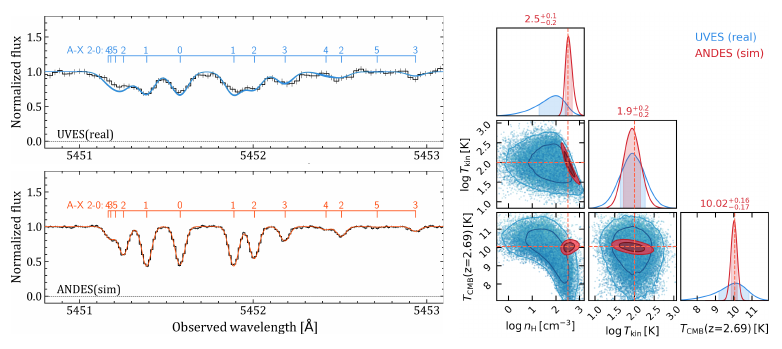}
\end{tabular}
\end{center}
\caption{ \label{fig:TCMB} 
Left: An example of CO absorption band observed at z = 2.69 towards the quasar J1237+0647 with VLT/UVES (R = 50000, S/N$\sim$40/pixel), and as expected with ANDES with R = 100000, S/N$\sim$100 pixel/1. The numbered tick lines indicate the different rotational levels that compose the band. Right: Simultaneous constraints on CMB temperature, density and kinetic temperature using profile fitting of CO lines coupled with excitation model for CO rotational levels. The blue and red colours represent the constraints obtained using existing UVES spectrum and a forecast for ANDES, respectively. Figure from Martins at al.\cite{martins:2024}.}
\end{figure}

\textbf{Evolution of the Cosmic Microwave Background Temperature ($T_{CMB}$).}
The evolution of the CMB temperature as a function of redshift is a fundamental prediction of the standard cosmological model. Deviations from the expected $T_{CMB}$(z) = $T_{CMB}$(0)(1 + z) relationship could signal new physics, such as violations of the Einstein Equivalence Principle or variations in the fine-structure constant. ANDES will leverage its high sensitivity and precision wavelength calibration to measure the CMB temperature at different redshifts with high accuracy. This capability is essential for testing the adiabatic expansion and photon conservation predictions of the standard model.
By detecting or placing stringent limits on deviations from the expected $T_{CMB}$ evolution, ANDES will explore potential new physical phenomena. The instrument's ability to achieve precise measurements of the thermal state of the IGM will be pivotal in identifying any such deviations (Fig. \ref{fig:TCMB}).

\begin{figure}[ht]
\begin{center}
\begin{tabular}{c} 
   \includegraphics[width=0.5\linewidth]{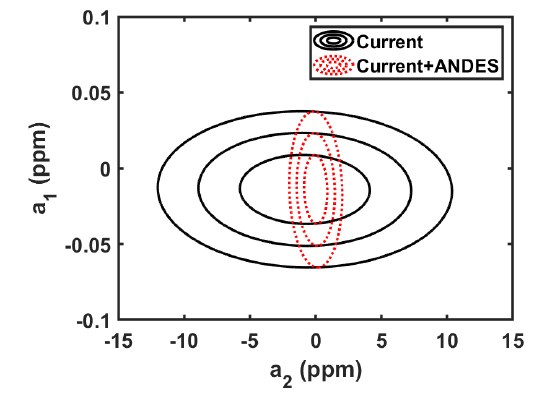}
\end{tabular}
\end{center}
\caption{ \label{fig:alpha} 
Model-independent constraints on the variation of the fine structure constant $\alpha$ parameterized as $\Delta\alpha/\alpha = a_1*y + (1/2)*a_2*y^2$  with $y = z/(1+z)$. $a_1$ is mainly constrained by laboratory experiments, while $a_2$ can only be constrained through astrophysics. Black contours show current constraints, while the red ones include a simulated dataset of ANDES measurements from 15 absorbers, with all other data unchanged. Figure from Martins at al.\cite{martins:2024}.}
\end{figure}

\textbf{Tests of the Universality of Physical Laws.}
One of the flagship science cases for ANDES involves testing the constancy of fundamental physical constants, such as the fine-structure constant ($\alpha$) and the proton-to-electron mass ratio ($\mu$). Any variation in these constants over cosmic time could indicate new forces or fields not accounted for in the Standard Model. Utilizing its extremely high spectral resolution, ANDES will conduct precise measurements of $\alpha$ and $\mu$ in various environments and redshifts, examining their constancy over time. This will involve comparing the relative frequencies of multiple transitions in quasar spectra, enabled by the instrument's broad wavelength range.
By investigating potential deviations from the universality of physical laws, ANDES will search for new interactions or particles. The ability to measure fundamental constants in different regions of space-time will help identify any composition-dependent forces, providing insights into the underlying physics.

\textbf{Real-Time Mapping of the Expansion History (Redshift Drift)}
ANDES will perform real-time measurements of the redshift drift, providing a direct, model-independent probe of the universe’s expansion history. This method, also known as the Sandage-Loeb test, measures the change in redshift of distant objects over time, offering a unique perspective on the dynamics of cosmic expansion.
ANDES will utilize its long-term stability to detect the subtle changes in redshift over decades, providing a direct and model-independent measurement of the cosmic expansion history. This will involve monitoring the redshifts of Ly$\alpha$ forest and other absorption lines in quasar spectra.
By combining redshift drift data with other cosmological observations, ANDES will constrain or refine models of dark energy and the overall cosmological paradigm. The instrument's high collecting power is essential for gathering sufficient photons from distant, faint quasars, achieving the precision required for this groundbreaking measurement.

Summarizing, ANDES at the ELT will significantly advance our understanding of cosmology and fundamental physics. By integrating high spectral resolution, broad wavelength coverage, and a large collecting area into its design, ANDES will enable detailed studies of the early universe, the evolution of fundamental constants, and the real-time expansion history of the cosmos. These capabilities position ANDES as a critical tool for exploring new physics and refining our cosmological models in the coming decades.
A detailed description of the science objectives for cosmology and fundamental physics is provided in the white paper by Martins at al.\cite{martins:2024}.

\subsection{Science Priorities}

Only a very expensive instrument would fulfil the requirements associated with the top priority science cases identified within each science area. Therefore, an overall science prioritization was performed by the core SAT during Phase A to drive the corresponding process of instrument design and to establish a trade-off between cost and scientific priorities. The following criteria were identified:
\begin{itemize}
\item Scientific impact: transformational versus incremental. 
\item Feasibility.
\item Competitiveness.
\end{itemize}
Also, if the TLR's of the top priority science case were enabling other science cases, the latter were not considered any further in the subsequent prioritization, as considered accomplished together with the top priority science case. The top science priorities are listed below and were used to define the technical specifications. We remark that these are not absolute science priorities, but science priorities identified with the aim of driving the instrument design.

 \begin{figure}[ht]
\begin{center}
\begin{tabular}{c} 
   \includegraphics[width=0.9\linewidth]{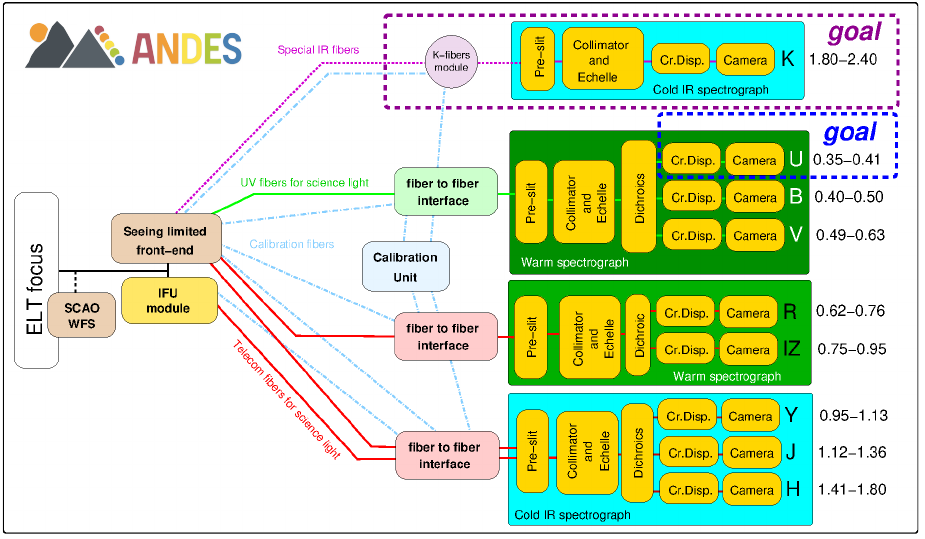}
\end{tabular}
\end{center}
\caption{\label{fig:baseline}ANDES architectural design, outlining the instrument subsystems: Front End (seeing-limited and AO assisted with SCAO unit), Fibre Link, Calibration Unit, [U]BV, RIZ, YJH and K (cold spectrographs).  
Andes Logo by Alexis Lavail (Uppsala).
}
\end{figure}

\begin{enumerate}

\item \textbf{Exoplanet atmospheres in transmission.} The TLRs for the study of exoplanet atmospheres (mainly $R\sim 100,000$, 0.5-1.8 $\mu m$ spectral range and wavelength calibration accuracy of 1 m/s) also enable the following science cases: 
reionization of the universe, the characterization of cool stars, the detection and investigation of near pristine gas, the study of extragalactic transients.

\item \textbf{Variation of the Fundamental Constants of Physics,} requiring an extension to 0.37 $\mu$m and automatically enabling to investigate:
the cosmic variation of the CMB temperature, the determination of the deuterium abundance, the investigation and characterization of primitive stars. 
At $\lambda < 0.40$mm the throughput of the ELT is expected to be low as a consequence of the planned coating. However, even in the range 0.37-0.40 $\mu$m the system is expected to be competitive with ESPRESSO. 

\item \textbf{Detection of exoplanet atmospheres in reﬂection,} requiring an Adaptive Optics (SCAO) system and an Integral Field Unit. These additional TLRs also automatically enable the following cases:  Planet formation in protoplanetary disks, Characterization of stellar atmospheres,  Search of low mass Black Holes. 

\item \textbf{Sandage test.} Its additional TLR of a stability of 1 cm/s, enables also: radial velocity searches and mass determinations of Earth-like exoplanets.

\end{enumerate}

This prioritization and the corresponding top level requirements finally resulted in the Technical Specifications whice were officially issued by ESO prior to the beginning of Phase B1 and which were used to drive the baseline design.

\section{Instrument concept}\label{sec:concept}
 \begin{figure}[ht!]
\begin{center}
\begin{tabular}{c} 
   \includegraphics[width=0.9\linewidth]{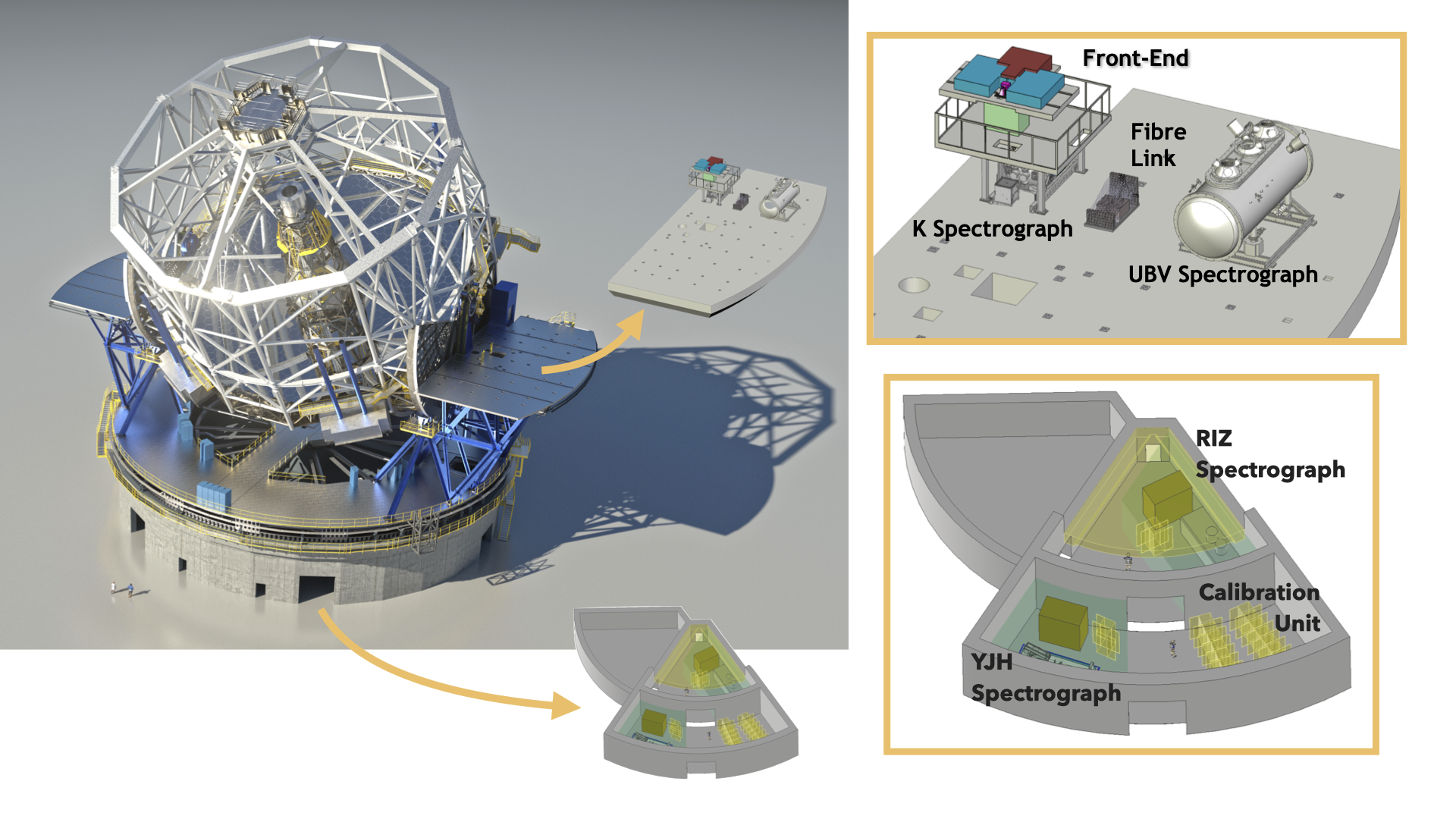}
\end{tabular}
\end{center}
\caption{\label{fig:coude}Overall and detailed views of ANDES Nasmyth and Coud\'e parts in the case the instrument is not entirely placed on the Nasmyth platform.}
\end{figure}
The ANDES baseline design is that of a modular fibre-fed cross dispersed echelle spectrograph which has three ultra-stable spectrometers, [U]BV, RIZ and YJH, providing a simultaneous spectral range of 0.4-1.8 $\mu$m at a resolution of $\sim$100,000. The goal is to extend the wavelength range to 0.35-2.4 $\mu$m with the addition of an U arm to the BV spectrograph and a separate K spectrograph.
The instrument can operate in seeing limited as well as in a SCAO-assisted mode. In seeing-limited mode it allows for simultaneous sky and/or calibration measurements. In SCAO mode it  uses a small IFU with at least two selectable spaxel scales (only in the YJH spectrometer, also in the RIZ spectrometer as a goal). The seeing-Limited mode will be quite unique among the ELT suite of instruments and will allow observations even in bad atmospheric conditions and/or with poor phasing of the ELT mirrors. It will also require simpler telescope operations.
The expected limiting magnitude for seeing limited observations is $m_{AB}$ = 20 in 1 hr with SNR=10 per resolution element. Simulations can be performed with the live ETC maintained by INAF-Arcetri at \href{https://andes.inaf.it/instrument/exposure-time-calculator/}{andes.inaf.it}.  This ETC can compute the limiting magnitude achievable at a given wavelength, for a given exposure time and at a given signal to noise ratio or it can compute the signal to noise ratio achievable at a given wavelength, in a given exposure time and at a given magnitude. 
With these observing modes and the expected performances, the proposed baseline design can fulfil the requirements of the 3 top science cases discussed in Sec. 3, with the goal of fulfilling also the 4th one. 
The proposed baseline of the system is shown in Fig. \ref{fig:baseline}, which provides a schematic view of the functional architecture and illustrates the modularity level chosen for the instrument. The major subsystems composing the instrument are presented as boxes, connected by fibers or fiber bundles that serves for delivering the light from the ELT focus and from the Calibration Unit(s) to the various Spectrographs that are located in different parts of the Telescope Infrastructure, the Nasmyth Platform and the Coud\'e room.  If enough volume and mass are available, the instrument can be placed entirely on the Nasmyth platform. Alternatively, the [U]BV and K modules can be placed on the Nasmyth platform while the other modules will sit in the Coud\'e Room (Figure \ref{fig:coude}).
The split in wavelengths over the modules is influenced, among all other parameters by the optical transparency of the different types of fibres available on the market. Therefore, the different modules can be positioned at different maximum distances from the telescope focal plane. The proposed configuration foresees that [U]BV is in the Nasmyth platform while the RIZ and YJH spectrometers are in the Coud\'e Room with the corresponding Fibre-to-Fibre interfaces and dedicated Calibration Units (that, for simplicity,  is represented as one single box in Figure \ref{fig:baseline}).
\begin{figure}[ht]
\begin{center}
\begin{tabular}{c} 
   \includegraphics[width=0.8\linewidth]{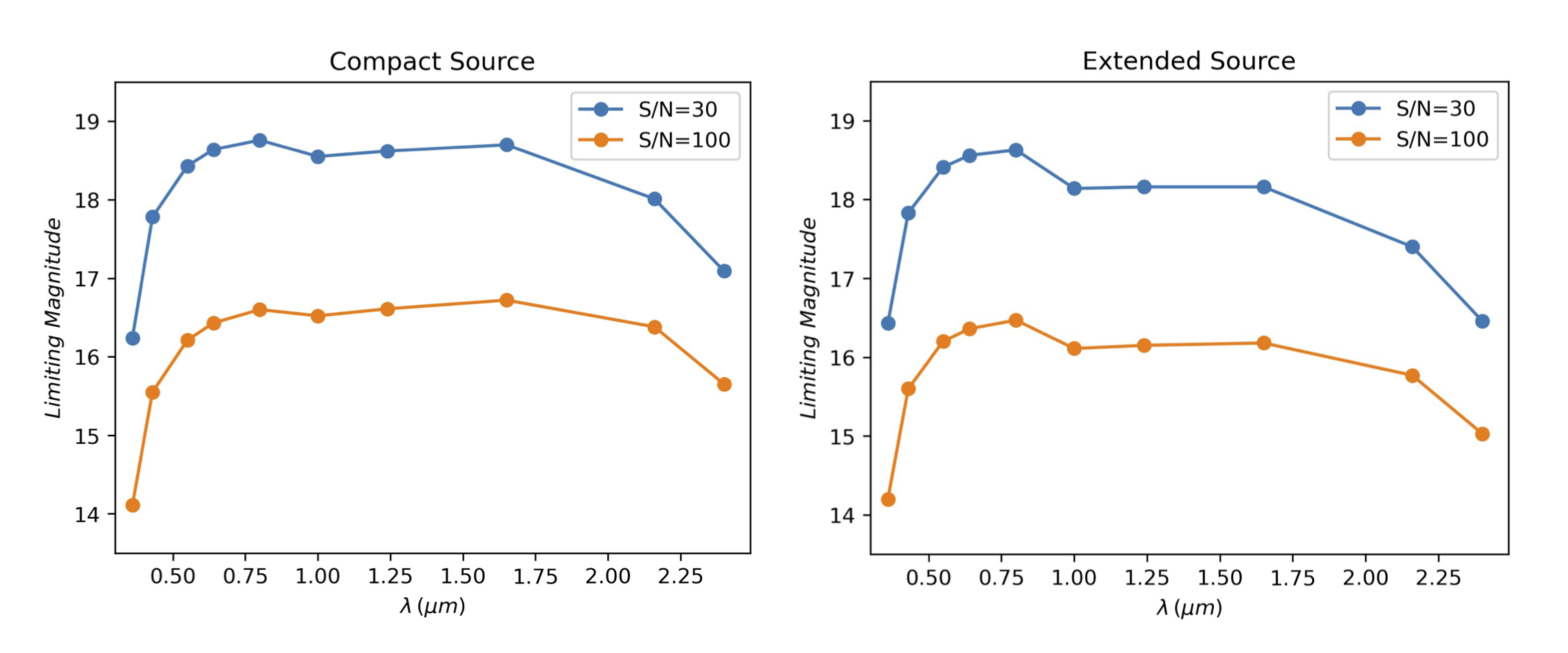}
\end{tabular}
\end{center}
\caption{\label{fig:perf} ANDES limiting magnitudes obtained from the ETC for different S/N ratios (30 – top and 100 – bottom), compact and extended sources (left and right).  Observations are in seeing-limited mode with R = 100,000 a total exposure time of 1800s. }
\end{figure}

During operations, the light from the telescope pre-focal station reaches the Front-End, a structure that is positioned on the Nasmyth platform and includes four independently insertable modules: two seeing limited arms, one SCAO arm and one IFU arm.

Depending on the chosen operational mode, the instrument is capable of inserting the different modules and delivering the light to specific modules.
Regardless of the specific observing mode, the Front End splits the light from the telescope via dichroics, provides proper optical correction and feeds the fibre bundles with the light divided into different spectral channels according to the wavelength bands reported in the picture and part of the TRS. 
The Fiber-Link subsystem is in charge of transporting the light to the spectrographs (also down to the Coud\'e room, Fig \ref{fig:coude}) and forming a series of parallel entrance slits, each consisting of an array of micro-lenses optically coupled to a fiber bundle. 
The fibre bundles are divided into two segments and have two different routings. 
The first segment connects the Front End to the Fibre-to-Fibre interface boxes, while the second segment connects the Fibre-to-Fibre interface boxes to the spectrographs. The two segments carrying the IFU light are directly connected, to maximize throughput; while the segments carrying the light from the seeing-limited apertures are coupled via a double scrambler optical system, to optimize precision and accuracy. The Fibre-to-Fibre boxes lie close to the spectrometers, i.e., the BV interface is on the Nasmyth platform while the infrared (RIZ and YJH) interfaces are in the Coud\'e room.
All spectrometer modules have a fixed configuration, i.e., no moving parts and provide proper spectral resolution and sampling in the requested wavelength range.
The calibration unit includes proper calibration sources for feeding the Front-End and the Fibre-to-Fibre interfaces (part of the FL) through a suitable routing of fibres.
Several analyses have been performed in order to define the best trade-off for the wavelength splitting and in particular, much effort has been dedicated to push the spectrographs as much as possible toward the goal ranges, including the U-Band in the BV spectrograph and an additional subsystem (fibres and spectrograph) for the K-band to be located on the Nasmyth platform. 

Summarizing, the main subsystem of ANDES are as follows (see also Figure \ref{fig:baseline}):
\begin{itemize}
\item Front End (FE) which transfers the light from the ELT Nasmith focus to the rest of the instrument, with 4 separated modules: two seeing limited and two for the IFU and SCAO modules. 
\item Fibre Link (FL) which couples the light coming from the Front End fibre bundles into the fibre bundles that feed the spectrographs.
\item Single Conjugated Adaptive Optics (SCAO) aimed to provide the adaptive optics correction to the IFU feeding the YJH spectrograph. 
\item Calibration Unit (CU) whose aim is to deliver calibration light for the ANDES spectrographs on demand and is connected via fibres to the Front End through the Fibre Links. 
\end{itemize}
and the Spectrographs:
\begin{itemize}
\item UV-optical Spectrograph ([U]BV). The [U]BV module has 3 cameras: V, B and U (goal) with one scientific detector each. 
The U arm/camera is an option under study to extend the wavelength coverage toward the blue.
\item Red Spectrograph (RIZ). The RIZ module has 2 cameras: R and IZ, with one scientific detector each. 
\item Infrared Spectrograph (YJH). The IR module has 3 cameras, Y, J and H, with one scientific detector each. 
\item K band Spectrograph (K). The K band module is an option under study has one camera, with one scientific detector. 
\end{itemize}
 All spectrographs have a fixed configuration, i.e. no moving parts, allowing to fulfil the requirements on stability. They include a series of parallel entrance slits consisting of linear micro-lens arrays each optically coupled to the fibre bundles. The split in wavelengths between the spectrographs is influenced, among other parameters by the optical throughput of the different types of fibres available on the market; therefore, the different modules can be positioned at different distances from the focal plane of the telescope (see Fig. \ref{fig:coude} for more details).
\begin{figure}[t]
\begin{center}
\begin{tabular}{c} 
 \includegraphics[width=0.7\linewidth]{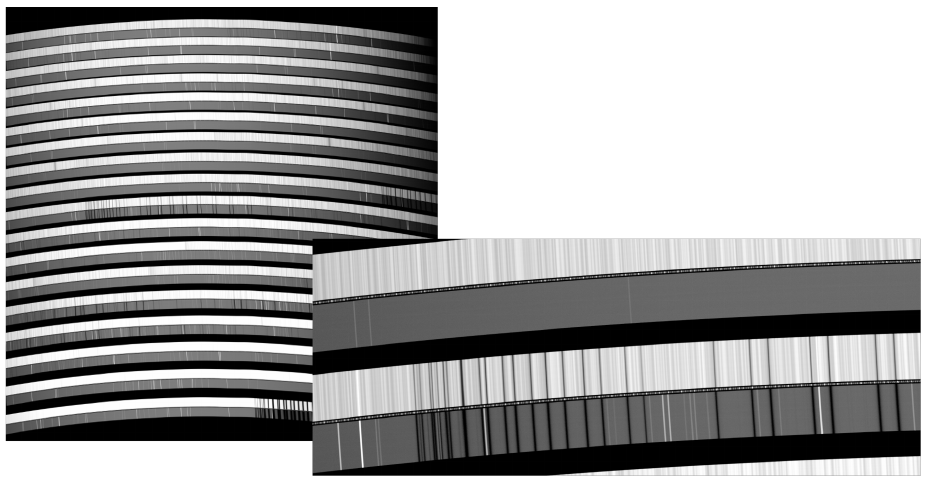}
\end{tabular}
\end{center}
\caption{\label{fig:e2e} Left: ANDES Simulated RIZ (R Band) science spectrum obtained with the E2E simulator.}
\end{figure}

The entire ANDES software (control and science tools) has been developed with the final goal of providing the astronomer with high level products as complete and precise as possible in a short time (within minutes) after the end of an observation while guaranteeing a smooth operation of the instrument at the Telescope when it will be offered to the community. This will maximize the overall efficiency and the scientific output making ANDES truly a “science-grade products generating machine” following the philosophy adopted already for ESPRESSO. 
Already in this phase ANDES-specific science tools have been developed: the ETC and the End-to-End simulator. 

The ETC is able to compute for a compact (extended) source the limiting magnitude (surface brightness) achievable at a given wavelength, in a given exposure time and at a given signal to noise ratio.  Moreover, this tool can compute the signal to noise ratio achievable at a given wavelength, in a given exposure time and at a given magnitude or surface brightness. Fig. \ref{fig:perf} shows the expected performance of ANDES in the cases of compact and extended sources. The ANDES ETC can be, at the time of writing, easily accessible at the address  \href{https://andes.inaf.it/instrument/exposure-time-calculator}{andes.inaf.it}.

The End-to-end (E2E) simulator has been used to perform full end-to-end simulations in order to evaluate the effect of technical choices on the science goals: an example is shown in Fig. \ref{fig:perf}, which shows a simulated ANDES raw data frame from the RIZ spectrograph (R band). 
E2E results coupled with the science requirements have also driven the identification of the main features of both DRS (Data Reduction Software) and DAS (Data Analysis Software), along with the Operation and Observation preparation tools. 

The ANDES DRS is the data reduction library for both visible and near-infrared parts of the spectrograph and it will be in charge to produce fully calibrated, science calibrated spectra starting from the RAW data secured by ANDES.  
The concept of DAS is instead modelled around the successfully experience of the ESPRESSO DAS as a set of self-standing modules ("recipes") performing interdependent operations. For both, a fully integrated approach based on ESO standard tools like CPL with ESORex and fully compliant with the ELT Dataflow.

The ANDES control software on the other side is responsible to control all the vital functions of the instrument like motors, sensors, lamps and to manage the execution of the scientific exposures according to the given observation mode. It will also implement all the needed interfaces towards external subsystems (telescope, scheduler, archive). During this phase a detailed requirement analysis have been conducted in order to identify main ANDES requirements leading to a quite detailed software architecture design and network architecture. The proposed architecture and solution for the current phase is fully based on the latest available version of ELT IFW framework. Moreover, also the main templates for ANDES observing mode, calibration and maintenance have been identified.

\section{The ANDES Consortium organization}

The ANDES Consortium is  composed of institutes from Brazil, Canada, Denmark, France, Germany, Italy, Poland, Portugal, Spain, Sweden, Switzerland, United Kingdom and USA. The full list of institutes is presented in Tab. \ref{tab:institutes}. Overall, the consortium includes 35 Institutes from 13 Countries. All Consortium members are listed as authors of this paper.

The large number of institutes involved in the Consortium reflects the large scientific and technical interest in a high-resolution spectrograph for the ELT in the astronomical community worldwide. In particular:
\begin{itemize}
\item the Consortium includes the large majority of the high-resolution community in the ESO member states and reflects the large scientific and technical interest in a high-resolution spectrograph for the ELT;
\item most Consortium Institutes have a proven expertise in building high-resolution spectrographs; overall the consortium includes most of the institutes that built the high-resolution spectrographs for ESO telescopes;
\item the Consortium can count on a large number of specialized technical and scientific staff as well as on world class laboratories and technical facilities.
\end{itemize}

\begin{table}[t]
\caption{\label{tab:institutes} Consortium Partners and Institutes}
\begin{center}
\begin{tabular}{|p{0.2\linewidth}|p{0.8\linewidth}|}
\hline
\textbf{Country} & \textbf{Consortium Partners} \\
\hline
\textbf{Brazil} & Board of Stellar Observational Astronomy, Federal University of Rio Grande do Norte, Natal \\
\hline
\textbf{Canada} & Observatoire du Mont-M\'egantic and the Institute for Research on Exoplanets, Universit\'e de Montr\'eal \newline
$~~-~$COPL (Centre for Optics, Photonic and Laser), University Laval\newline
$~~-~$Department of Physics \& Astronomy, University of Victoria\newline
$~~-~$Dunlap Institute + Dept. Astronomy \& Astrophysics, University of Toronto\newline
$~~-~$Quebec Artificial Intelligence Institute (Mila)\\
\hline
\textbf{Denmark} & Instrument Centre for Danish Astrophysics representing:\newline
$~~-~$Niels Bohr Institute, København\newline
$~~-~$Aarhus University\newline
$~~-~$Danmarks Tekniske Universitet (DTU), Lyngby
\\
\hline
\textbf{France} & Centre National de la Recherche Scientifique (CNRS) representing:\newline
$~~-~$LAGRANGE, Observatoire de la Côte d’Azur, Nice\newline
$~~-~$LAM (Laboratoire d’Astrophysique de Marseille), Marseille\newline
$~~-~$IRAP (Institut de Recherche en Astrophysique et Planetologie), Un. Toulouse\newline
$~~-~$IPAG (Institut de Plan\'etologie et d’Astrophysique), Un. Grenoble Alpes\newline
$~~-~$LUPM (Laboratoire Univers et Particules), Universit\'e de Montpellier\newline
$~~-~$IAP (Institut d’Astrophysique de Paris)\newline
$~~-~$LMD (Laboratoire de M\'et\'eorologie Dynamique), Ecole Polytechnique\\
\hline
\textbf{Germany} & Leibniz-Institut für Astrophysik Potsdam (AIP)  \\
& Institut für Astrophysik und Geophysik, Universität Göttingen (IAG) \\
& Max-Planck-Institut für Astronomie, Heidelberg \\
& Zentrum für Astronomie (ZAH), Universität Heidelberg  \\
 & Thüringer Landesternwarte Tautenburg (TLS) \\
 &Hamburger Sternwarte, Universit\"t Hamburg (UHH)\\
\hline
\textbf{Italy} & Istituto Nazionale di Astrofisica (INAF), 'Leading Technical Institute'\\
\hline
\textbf{Poland} & Nicolaus Copernicus University in Toruń \\
\hline
\textbf{Portugal} & Instituto de Astrof\'isica e Ciências do Espaço, Porto\\
& Centro de Investigação em Astronomia/Astrof\'isica da Universidade do Porto\\
& Associação para a Investigação e Desenvolvimento de Ciências, Universidade de Lisboa \\

\hline
\textbf{Spain} & Instituto de Astrof\'isica de Canarias  \\
& Consejo Superior de Investigaciones Cient\'ificas (CSIC, Spain) representing:\newline
$~~-~$Instituto de Astrof\'isica de Andaluc\'ia (IAA)\newline
$~~-~$Centro de Astrobiolog\'ia (CSIC-INTA) \\
\hline
\textbf{Sweden}& Lund University\\
& Stockholm University\\
& Uppsala University\\
\hline
\textbf{Switzerland}& D\'epartement d’Astronomie, Universit\'e de Genève\\
 & Physikalisches Institut, Universität Bern \\
\hline
\textbf{United Kingdom} & Science and Technology Facilities Council representing:\newline
$~~-~$UK Astronomy Technology Centre\newline
$~~-~$Cavendish Laboratory \& Institute of Astronomy, University of Cambridge\newline
$~~-~$Institute of Photonics and Quantum Sciences, Heriot-Watt University  \\
\hline
\textbf{USA} & Department of Astronomy, University of Michigan\\
\hline
\end{tabular}
\end{center}
\end{table}

\begin{figure} [t]
\begin{center}
\begin{tabular}{c} 
   \includegraphics[width=0.9\linewidth]{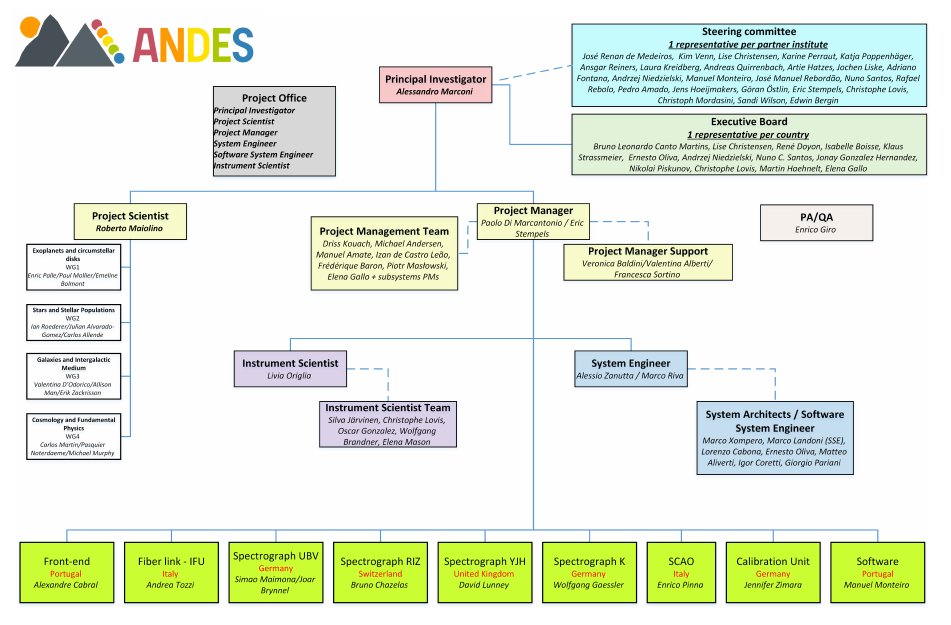}
\end{tabular}
\end{center}
\caption[example]{\label{fig:organization} ANDES Consortium organization breakdown structure. 
ANDES logo by Alexis Lavail (Uppsala)
}
\end{figure}

The ANDES Organization Breakdown Structure and the names of the key persons in the project are shown in Fig. \ref{fig:organization}.

The consortium is represented by the Principal Investigator who has the ultimate responsibility of the project and is the formal contact point between ESO and the Consortium. 
The Steering Committee (SC) is the ultimate decision-making body of the Consortium and the source of guidance and strategic decision making, in particular it approves all decisions significantly impacting the financial and human resources allocated to the Project. It is composed of one representative per Partner. 
The Executive Board (EB) provides regular advice to the PI and the SC on all technical and scientific matters, in order to ensure the fulfilment of the scientific objectives of the Project. It is a strong but hands-off high-level overview committee. Together with the PI prepares proposals for the decisions of the SC. 
With a loose analogy we could associate the PI to the Prime Minister, the EB to the Government, and the SC to the Parliament.

The Project Scientist leads the development of the science program and top-level requirements, working in close contact with the Project Manager (and his team) who takes the responsibility to manage the overall project. The project manager is supported by a System Engineer and the system team (composed by the Software System Engineer and System Architects) who are in charge to supervise the overall system design. SE and SSE work in close contact with the Instrument Scientist (and her team) who makes sure that the adopted technical solutions match the foreseen instrument scientific needs. The PI, PM, SE, SSE, PS and IS form the Project Office which ultimately and collegially encompass all programmatic activities of the project.

Technical work is organized following a modular approach where ANDES is seen as composed by the 8 subsystems described in Sec. \ref{sec:concept}: Front end, Fiber Link – Integral Field Unit, [U]BV spectrograph, RIZ spectrograph, YJH spectrograph, K spectrograph, Single Conjugated Adaptive Optics module, Calibration Unit, Software.
For each subsystem, a subsystem Project Manager and a subsystem System Engineer have been appointed with similar responsibilities of the PM and SE even though limited to the subsystem level.

All matters that concerns science cases and GTO scientific programme for ANDES are responsibility of the PS and the Science Team (ST). The ST is organized in four Working Groups (WGs):  WG1- Exoplanets and Circumstellar disks, WG2 - Stars and Stellar populations, WG3 - Galaxies (formation and evolution) and Intergalactic Medium, WG4 - Cosmology and Fundamental Physics.
Each WG is led by a chair, assisted by co-chair(s) and composed by a number of scientists from the participating countries selected by appointment from the PI, following indications of the EB and of the PS and ultimately approved by the SC.

The Consortium is assisted by a follow-up team from ESO, composed of a Project Manager, an Instrument Scientist and a System engineer which provide all the necessary information and support to design and build an instrument which is compliant with ELT standards and with the scientific needs of the ESO community.

Communication and Outreach. The ANDES Consortium recognizes the important of Communication and Outreach both internally and externally to the Consortium. This is why a Communication and Outreach WG has been established which includes at least one representative per major subsystem/country. In this intial phase sof the project, the WG activity is focused to plan an internal communication, while in a more advanced stage it will develop different strategies to communicate results and milestones to external stakeholders.
The website has a crucial role in this strategy and for that the WG has recently restyled the webpage (\href{https://andes.inaf.it/}{andes.inaf.it}) where ANDES is presented in its fundamental outlines: scientific cases, technical design, consortium and team, followed by useful tools (internal documents, exposure time calculator, access to restricted clouds, and so on), publications and conferences, the news section.

\section{Cost Estimates and schedule}

The current estimate of the overall hardware costs is around 45 MEUR in hardware (including the K band and excluding contingencies). Contingencies have been evaluated at subsystem level using different margins according to the considered components (usually 5-10\%). Then a 5\% contingency at system level can been applied. We have considered a contingency lower than 20\% because the proposed baseline design is based on proven technical solutions and can benefit on heritage from ESPRESSO and other similar instruments developed by Consortium Partners.
The FTEs required to complete the project are estimated, to be approximately 650 when including contingencies.
The FTEs will be compensated with 65 nights of GTO, while the hardware costs will be compensated with a number of GTO nights still to be agreed, but with a minimum of at least 60. These GTO nights will be used for joint Consortium science programs.

The current Project schedule foresees that ANDES will be available at the telescope in 2032, but the we could adopt a more aggressive schedule to send a module at the telescope as early as 2030/2031, if needed. Indeed, a more aggressive schedule could be adopted to minimize the risk that the key ANDES science cases are harvested by high-resolution spectrographs at other 30m-class telescopes (e.g., G-CLEF at GMT).
It is worth mentioning that the Consortium will do  subsystem reviews in October/November 2024, in order to have a PDR in June 2025 and minimize delays. 

\section{Conclusions}

The ANDES baseline design is that of three ultra-stable and modular ﬁbre-fed cross dispersed echelle spectrographs providing a simultaneous spectral coverage of 0.4-1.8 $\mu$m (goal 0.35-2.4 $\mu$m) at a resolution of 100,000 with several, interchangeable, observing modes ensuring maximization of either accuracy, throughput or spatially resolved information. Overall, the studies conducted so far have shown that the ANDES baseline design can address the 4 top priority science cases, being able to provide ground-breaking science results with no obvious technical showstoppers. 

The construction of ANDES includes the majority of the institutes in ESO member states with expertise in high resolution spectroscopy and will require an estimated 45 MEUR in hardware (including the K band and excluding contingencies) and about 650 FTEs. Contingencies are expected to be low (5-10\%) because the proposed baseline design is based on proven technical solutions and can beneﬁt on heritage from HARPS and ESPRESSO and other previous high-resolution spectrographs, e.g. PEPSI at the 11.8m LBT, SPIRou, CARMENES and GIANO. The construction (including Phase B) will last about 8-10 years. Therefore, with Phase B concluding in 2025, ANDES (or at least one of its modules) could be at the telescope in 2031/2032.

Overall, ANDES is an instrument capable of addressing ground-breaking science cases while being almost (telescope) pupil independent, as it can operate both in seeing and diffraction limited modes; the modularity ensures flexibility during construction and the possibility to quickly adapt to new development in the technical as well as science landscape.

We conclude by summarizing  the highlights of the ANDES project, formerly known as HIRES, from Phase A through Phase-B1.
\begin{itemize}
\item The Phase A study has been conducted by an international consortium including the majority of the institutes in ESO member states expert in high resolution spectroscopy. During Phase A the consortium was composed of 30 institutes from 12 countries, and a grand total of about 200 people have contributed to the study. The Consortium is now composed of 33 institutes in 13 countries, with almost 300 people contributing to the project.
\item The Phase A study started on March 22, 2016 and has been completed in one and a half years, with the datapack delivery to ESO on October 6, 2017.
\item The ANDES Construction has been officially approved by the ESO Council in December 2021.
\item Phase-B1  started in September 2022 and was concluded with the System Architecture Review (SAR) in October 2023.
\item The science priorities which have been identified for prioritizing the Top Level Requirements of the instrument are:
\begin{enumerate}
\item the detection of the signatures of life through the study of exoplanet atmospheres in transmission
\item the variation of the fundamental constants of Physics
\item the detection of the signatures of life through the study of exoplanet atmospheres in reflection
\item the direct detection of the Cosmic acceleration through the measurement of the Sandage effect
\end{enumerate}
Many other ground-breaking science cases are made possible with the TLRs inferred from 1 to 4. These TLRs has then drive the Technical Specification which has been officially issued by ESO.

\item The ANDES baseline design is that of a modular fibre-fed cross dispersed echelle spectrograph which has three ultra-stable spectral arms, [U]BV, RIZ and YJH, providing a simultaneous spectral range of 0.4-18 $\mu$m (goal 0.35-2.4 $\mu$m with an extension to the U band and additional spectral arm K) at a resolution of 100,000 with several, interchangeable, observing modes allowing observations both  in seeing- and diffraction-limited modes.
\item In particular, the seeing-limited mode will be quite unique among the ELT suite of instruments and will allow observations even in bad atmospheric conditions and/or with poor phasing of the ELT mirrors. It will also require simpler telescope operations.
\item The proposed baseline design can fulfil the requirements of the 3 top science cases, with the goal of fulfilling also the 4th one.
\item The modularity of the instrument ensures that several extensions can be added without affecting the instrument design or stability. The fibre-feeding ensures ultra-stable VIS and NIR spectrographs with no internal moving parts while, at the same time, allowing for many different observing modes.
\item The total estimated cost of the baseline design is about 45 MEUR, excluding contingencies at the level of $\sim 10\%$, with an estimated 650 FTEs required to complete the project.
\item The consortium will provide all the necessary funds for the construction of the instrument. ESO will compensate the consortium for  hardware and FTEs costs with GTO on the ELT, which the Consortium will use for joinst science programs.
\item The schedule proposed by the Consortium now foresees PDR in June 2025 with the final Preliminary acceptance in Chile in 2032, part of which could be anticipated to 2030/2031 if needed.
\item Overall, ANDES should start operating as close to the ELT first light as possible and the studies conducted so far have provided several reasons for this:
\begin{enumerate}
\item It allows ground-breaking science cases, and it should not be late compared to its competitors
\item As an ELT instrument, it is relatively technically simple and has relatively low risks
\item It is almost independent of the quality of the telescope pupil and it can operate in seeing-limited conditions
\item It is a modular instrument and therefore it naturally allows a staged deployment.
\end{enumerate}
\end{itemize}

\acknowledgments 
 
The Italian effort for ANDES is supported by the Italian National Institute for Astrophysics (INAF).\\
AIP received financial support by the German Federal Ministry of Education and Research (BMBF/DESY-PH: 05A23BAB)

The Portuguese team thanks the Portuguese Space Agency for the provision of financial support in the framework of the PRODEX Programme of the European Space Agency (ESA) under contract number 4000143136 as well as from funds by the European Union (ERC, FIERCE, 101052347). Views and opinions expressed are however those of the author(s) only and do not necessarily reflect those of the European Union or the European Research Council. Neither the European Union nor the granting authority can be held responsible for them. This work further was supported by FCT - Fundação para a Ciência e a Tecnologia through national funds and by FEDER through COMPETE2020 - Programa Operacional Competitividade e Internacionalização by these grants: UIDB/04434/2020; UIDP/04434/2020.
The ANDES project is partially funded through the SNSF FLARE programme for large infrastructures under grants 20FL21\_173604, 20FL20\_186177 and 20FL20\_216577
Swedish participation in the ANDES project is made possible through the national  Swedish ELT Instrumentation Consortium (SELTIC), suppored by the Swedish Research 
Council (VR).
%
%
CJM acknowledges FCT and POCH/FSE (EC) support through Investigador FCT Contract 2021.01214.CEECIND/CP1658/CT0001 and project
2022.04048.PTDC (Phi in the Sky, DOI 10.54499/2022.04048.PTDC)
NCS acknowledges funding by the European Union (ERC, FIERCE, 101052347). Views and opinions expressed are however those of the author(s) only and do not necessarily reflect those of the European Union or the European Research Council. Neither the European Union nor the granting authority can be held responsible for them. This work was supported by FCT - Fundação para a Ciência e a Tecnologia through national funds and by FEDER through COMPETE2020 - Programa Operacional Competitividade e Internacionalização by these grants: UIDB/04434/2020; UIDP/04434/2020.

JLB acknowledges funding from the European Research Council (ERC) under the European Union’s Horizon 2020 research and innovation program under grant agreement No 805445.
MTM acknowledges the support of the Australian Research Council through Future Fellowship grant FT180100194
SS acknowledges funding from the European Research Council (ERC) under the European Union’s Horizon 2020 research and innovation program under grant agreement No 804240.
TMS acknowledges support from the SNF synergia grant CRSII5-193689 (BLUVES). 
CAP, JIGH, RR, ASM, MA, FGT, JPC, FTS, AVM, and RS acknowledge financial support from the Spanish Ministry of Science and Innovation (MICINN) project PID2020-117493GB-I00. 
PJA, CRL, RCO, RV acknowledge financial support from the Agencia Estatal de Investigaci\'on (AEI/10.13039/501100011033) of the Ministerio de Ciencia e Innovaci\'on and the ERDF “A way of making Europe” through projects PID2022-137241NB-C43 and the Centre of Excellence “Severo Ochoa” award to the Instituto de Astrof\'isica de Andaluc\'ia (CEX2021-001131-S).
%
%
\bibliography{biblio} 
\bibliographystyle{spiebib} 

\end{document}